\documentclass[fleqn,twoside]{article}
\usepackage{espcrc2}

\usepackage{epsfig}
\usepackage[figuresright]{rotating}

\newcommand{\AmS}{{\protect\the\textfont2
  A\kern-.1667em\lower.5ex\hbox{M}\kern-.125emS}}

\title{Minireview of Leptoquark Searches}

\author{P.~Bruce Straub
  \address[MCSD]{Yale University\\
        DESY, F1/ZEUS\\ 
        Notkestrasse 85\\
        22607 Hamburg, Germany}}%
       
\newcommand{\MLQ}{M_{LQ}}
\newcommand{\Mlj}{M_{\ell j}}
\newcommand{\alphaem}{\alpha_{\rm em}}
\newcommand{\lambdaem}{\sqrt{4\pi\alphaem}}
\newcommand{\Figbb}[4]{\epsfig{figure=plots/#1,bb=#4,width=#2cm,height=#3cm,clip=}}
\newcommand{\cFigbb}[4]
{\begin{center}\Figbb{#1}{#2}{#3}{#4}\end{center}}
\begin{document}

\begin{abstract}
Direct and indirect leptoquark searches at colliders are reviewed.
\vspace{1pc}
\end{abstract}

\maketitle

\section{Introduction}
Leptoquarks are hypothetical bosons which couple to a lepton and
a quark via a Yukawa coupling (denoted $\lambda$). In the Standard model,
both quarks and leptons occur in left-handed $SU(2)$ doublets and
right-handed $SU(2)$ singlets. The symmetry between quarks and leptons
leads to the cancellation of triangle anomalies which make the SM
renormalizable. Leptoquarks appear in theories\cite{theories} in which
this symmetry is more fundamental.

Leptoquarks (LQs) are color triplets, which would be pair produced in
either $q\bar{q}$ or $gg$ interactions at $p\bar{p}$ or $pp$ colliders.
Because they carry electroweak charge, they would also be pair produced
in $e^+e^-$ or $\mu^+\mu^-$ collisions.
Only standard model gauge couplings are involved in pair production;
therefore the cross sections depend neither on the
quark-lepton-LQ Yukawa coupling nor on the quark and lepton generations
to which the leptoquark couples.
In contrast, leptoquarks would be singly produced via the Yukawa coupling
in a lepton-quark collision. Searches at electron-proton colliders are
sensitive only to LQs which couple to electrons and the
sensitivity to LQs which couple to second and third generation quarks is
far below that of first-generation LQs.
Leptoquarks are usually (but not always) assumed to be {\it generation
diagonal}.  Models in which LQs couple to more than one generation of
quarks or leptons would induce flavor-changing neutral
currents\cite{Davidson,Leurer} or
lepton flavor violation\cite{Davidson,Gabrielli} respectively.

The model of Buchm{\"u}ller, R{\"u}ckl and Wyler\cite{BRW} (BRW), in which
leptoquarks couple to a single generation of SM fermions via 
chiral Yukawa couplings which are invariant under $SU(3)\times
SU(2)\times U(1)$ is often used to classify possible leptoquark
species.  In the BRW model baryon and lepton numbers ($B$ and $L$) are
conserved; there exist ten possible leptoquark species
characterized by the chirality of the coupling, the the spin ($J=0$ or 1),
the weak isospin ($T=$0, $1/2$, or 1), and the fermion number,
$F=3B+L=0$ or $2$. 
Four of the species could have separate couplings to right- and left-handed
leptons. A leptoquark which coupled first generation quarks
to both left- and right-handed electrons would
mediate $\pi\to e\nu$ decay\cite{BW}. The agreement of the measured decay rate
with SM predictions motivates the conventional assumption that only one
of the couplings can be nonzero. 
In the most commonly used notation\cite{AachenNotation}
$F=0$ scalar ($S$) and vector ($V$) species are denoted 
$S^L_{1/2}$, $S^R_{1/2}$, ${\tilde S}^L_{1/2}$, 
$V^L_0$, $V^R_0$, ${\tilde V}^R_0$ and $V^L_1$, while
the $F=2$ species are $S^L_0$, $S^R_0$, ${\tilde S}^R_0$, $S^L_1$,
$V^L_{1/2}$, $V^R_{1/2}$, and ${\tilde V}^L_{1/2}$.
The sub- and superscripts indicate the lepton chirality and the weak isospin
respectively. The species ${\tilde S}$ and ${\tilde V}$ differ
by two units of hypercharge from $S$ and $V$ respectively.
Partial widths for scalar and vector LQs with mass $\MLQ$
are given by $\lambda^2\MLQ/16\pi$ and $\lambda^2\MLQ/24\pi$ respectively.

\section{Leptoquark Searches the Tevatron}
The Tevatron experiments have performed searches
for first, second, and third generation leptoquarks, including separate
analyses for decays to charged leptons and to neutrinos. The results
are summarized in table \ref{tab:Tevatron}. 
Limits are more restrictive for LQs which decay to $eq$ and $\mu q$
than those decaying to $\nu q$. CDF has used their silicon vertex detector
to tag $b$ and $c$ quarks\cite{CDFnunubb} while D0 identifies $b$-jets
by associated muons\cite{DZeronunubb}.
For scalar LQs, NLO cross section calculations\cite{ppscalarcs} are used
in limit setting. In vector LQ cross section calculations\cite{ppvectorcs}
(done at LO), the coupling of the LQ and gluon fields is model dependent
and limits for several different models\cite{DZeroFirstGen,CDFtautaubb}
are quoted.

\begin{table}\begin{tabular}{crrrl}
     & &\multicolumn{2}{c}{$\MLQ$ Limit (GeV)}&\\
 Gen.& $\beta$&Scalar&Vector& Ref.\\ \hline
   1 & 1     & 225 & 246,345 & D0\cite{DZeroFirstGen,DZeroeeqq,DZeroenuqq}\\
     & $1/2$ & 204 & 233,337 & \\  \hline
   1 & 1     & 213 &         & CDF\cite{CDFeeqq} \\ \hline
   1 & 1     & 242 &         & CDF+D0\cite{CDFDZero} \\ \hline
   2 & 1     & 200 & 275,325 & D0\cite{DZeromumuqq,DZeromunuqq} \\
     & $1/2$ & 180 & 260,310 & \\ \hline
   2 & 1     & 202 &         & CDF\cite{CDFmumuqq,CDFnunubb} \\
     & $1/2$ & 160 &         & \\ 
     & 0     & 123 & 171,222 & \\ \hline
 1,2 & $0$   &  98 & 200,298 & D0\cite{DZeronunuqq} \\ \hline
   3 & 0     &  94 & 148,209 & D0\cite{DZeronunubb} \\ \hline
   3 & 1     &  99 & 170,217 & CDF\cite{CDFtautaubb,CDFnunubb} \\ 
     & 0     & 148 & 199,250 & \\ \hline
\end{tabular}
\caption{Limits on leptoquarks  from
the Tevatron. The first two columns give the generation
and $\beta$, the branching ratio to a charged lepton
and a quark. The next two columns give the 
lower bounds on $\MLQ$ (95\% CL). Vector LQs limits are given
for both low and high cross section models.}\label{tab:Tevatron}
\end{table}
\section{Leptoquark Searches at HERA}
\begin{figure}
\cFigbb{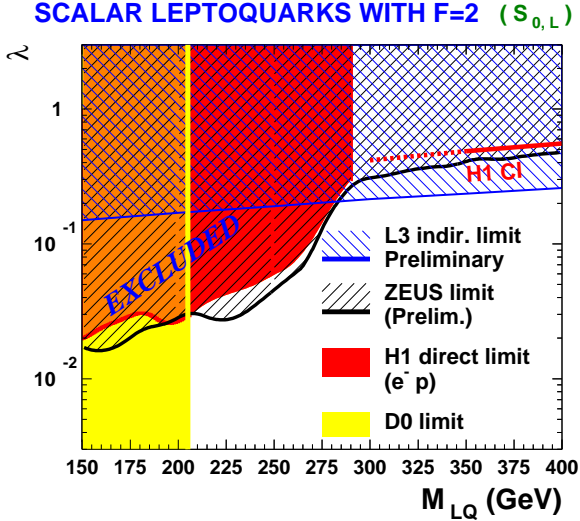}{6}{6}{80 205 480 600}
\caption{Limits on the $S_0^L$ leptoquark
in the $(\MLQ,\lambda)$ plane.
The excluded region lies 
above the curves for H1, L3, and ZEUS
and to the left of the line for D0.}
\label{fig:s0L}
\end{figure}

At the HERA electron-proton collider, first-generation
leptoquarks can be singly produced via fusion of the beam electron
with valence quarks in the proton. Event by event,
the electron-jet and neutrino-jet final states resulting from
LQ production (or exchange of virtual LQs) are indistinguishable
from Neutral and Charged Current Deep Inelastic Scattering (DIS).
If $\MLQ$ is below the center of mass energy,
($\sqrt{s}=318$\,GeV) and the width of the LQ state is small, LQs would
be produced as an $s$-channel resonance with cross section proportional
to $\lambda^2$. Leptoquark production and decay would give rise to a
a peak in the spectrum of lepton-jet invariant mass ($\Mlj$), while the
DIS background appears as a continuum.
The lepton-jet invariant mass is related
to the fraction $x$ of the proton momentum carried by the interacting quark
by $\Mlj^2=xs$.
A LQ signal is distinguished from the DIS background by the angular
distribution of the final-state leptons. Define $\theta^*$ as
the angle between the incident electron and the final state lepton in
the LQ rest frame and $y=1-\cos\theta^*$.
The distribution of $y$ for scalar LQ events is flat while
vector LQs have a $1-y^2$ distribution. In contrast,
the NC-DIS background follows a $1/y^2$ distribution (for $\gamma$ exchange).
Both H1\cite{HoneDirect,HoneDirectLFV} and ZEUS\cite{ZEUSDirect}
have searched for LQs by computing a
$\lambda$-dependent likelihood for the 2-dimensional
distributions of electron-jet and neutrino-jet events in $\Mlj$ and
$y$ (or $\cos\theta^*$). This method is also sensitive to the
changes in the observed $(\Mlj,\cos\theta^*)$ distribution resulting from
$u$- and $s$-channel LQ diagrams with
$\MLQ$ \raisebox{-1.1mm}{$\stackrel{>}{\sim}$} $\sqrt{s}$.
Example of the resulting limit curves are
shown in Fig.~\ref{fig:s0L}.
In table \ref{tab:LEPHERA}, the ZEUS lower limits on $\MLQ$ for 
$\lambda=0.1$ are shown. Both H1\cite{HoneIndirect} and ZEUS\cite{ZEUSIndirect}
have also obtained limits for LQs with $\MLQ>\sqrt{s}$ by examining the
$Q^2$ spectrum of NC-DIS-like events ($Q^2$ is the negative square of the
4-momentum transfer between the incident $e$ and the hadronic system).
Limits obtained in these analyses
are consistent with those in the analyses described above.
In addition ZEUS\cite{ZEUSLFV} and H1\cite{HoneDirectLFV} have searched by
LQs which couple to two different lepton generations, thereby inducing
lepton flavor violation.
\section{Leptoquark Searches at LEP}
\begin{figure}
\cFigbb{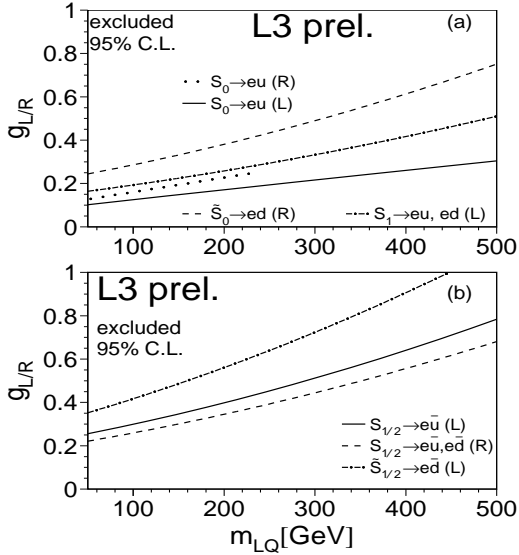}{6}{6}{60 120 557 680}
\caption{Scalar leptoquark limits 
in the  $(\MLQ,\lambda)$ plane from L3\cite{LLLindirect}.
The area above the curves is excluded.}
\label{fig:L3indirect}
\end{figure}

Three sorts of leptoquark searches have been performed at LEP.
Leptoquarks with $\MLQ<\sqrt{s}/2$ would be pair
produced via electroweak coupling. 
Using data with $\sqrt{s}$ from 189 to 209 GeV OPAL\cite{OPALpair}
searched for scalar and vector LQs from
all three generations, obtaining lower limits on $\MLQ$ ranging from
72 to 102 GeV. Single LQ production can occur via the interaction
of an electron with a radiated photon which resolves into
$q\bar{q}$. Limits in the $(\MLQ,\lambda)$ plane for
various LQ species obtained by OPAL\cite{OPALsingle}
range from 189 to 209 GeV for $\lambda=\lambdaem$,
where $\alphaem$ is the fine structure constant.
See \cite{DELPHIsingle} for DELPHI results.
The most interesting LQ searches\cite{LLLindirect,OPALindirect,ALEPHindirect}
performed at LEP look for indirect effects of $t$-channel ($F=0$) or
$u$-channel ($F=2$) LQ exchange in the reaction $e^+e^-\to q\bar{q}$.
L3 limits on
$\lambda$ vs. $\MLQ$ for scalar LQs are shown in Fig.\ \ref{fig:L3indirect}.
\begin{table}[ht]
\begin{tabular}{crrrr}
species & \multicolumn{3}{c}{LEP} & ZEUS \\
& 1$^{\rm st}$ & 2$^{\rm nd}$ & 3$^{\rm nd}$ & 1$^{\rm st}$\\ \hline
$S_0^L$              & 655 & 762 &  NA  & 274 \\
$S_0^R$              & 520 & 625 &  NA  & 272 \\
${\tilde S}_0^R$     & 202 & 209 &  454 & 249 \\
$S_{1/2}^L$          & 178 & 215 &  NA  & 281 \\
$S_{1/2}^R$          & 232 & 185 &    - & 281 \\
${\tilde S}_{1/2}^L$ &   - &   - &  226 & 270 \\
$S_1^L$              & 361 & 408 & 1036 & 274 \\ \hline
$V_0^L$              & 917 & 968 &  765 & 268 \\
$V_0^R$              & 165 & 147 &  167 & 271 \\
${\tilde V}_0^R$     & 489 & 478 &  NA  & 279 \\
$V_{1/2}^L$          & 303 & 165 &  208 & 248 \\
$V_{1/2}^R$          & 227 & 466 &  499 & 273 \\
${\tilde V}_{1/2}^L$ & 176 & 101 &  NA  & 272 \\
$V_1^L$              & 659 & 687 &  765 & 283 \\ \hline
\end{tabular}
\caption{Lower limits on $\MLQ$ in GeV for different LQ species
which couple to first-generation leptons. The first column gives the
LQ species; the next three columns give the LEP combined limits for
first, second, and third generation quarks with $\lambda=\lambdaem$.
A dash indicates that no limit could be set and NA indicates an LQ
which couples only to top quarks, which would not be visible at LEP.
The last column gives the ZEUS limits for $\lambda=0.1$.}
\label{tab:LEPHERA}
\end{table}
Table \ref{tab:LEPHERA} shows combined LEP limits\cite{LEPff}
on LQs which couple to $e$ and quarks of different generations.
These limits were obtained using the
cross section for $e^+e^-\to$ hadrons for $\sqrt{s}$ between 130 and 207 GeV.
For LQs which couple to $ec$ and $eb$, the
the $c\bar{c}$ and $b\bar{b}$ fractions $R_c$ and $R_b$, and
the forward-backward asymmetries 
were also used.
\section{Summary}
For first generation leptoquarks, the Tevatron experiments have set limits
on scalars (coupling to $eq$) of $\MLQ>242$ GeV and corresponding vector
LQ limits in the range from 233 to 345 GeV, depending on model assumptions.
The LEP experiments have set lower limits on $\MLQ$ (which are approximately
proportional to $\lambda$) which range from 165 to
917 GeV for $\lambda=\lambdaem\sim0.31$. The HERA experiments have set
lower mass limits in the range of $\sim250$ to 280 GeV for $\lambda=0.1$.
In addition, searches at the Tevatron and LEP have constrained LQs
coupling to leptons and quarks of the second and third generations.
\newcommand{\PRL}[4]{#1, Phys.\ Rev.\ Lett.\ {\bf #2}, #3 (#4)}
\newcommand{\PLB}[4]{#1, Phys.\ Lett.\ {\bf B#2}, #3 (#4)}
\newcommand{\PRD}[4]{#1, Phys.\ Rev.\ {\bf D#2}, #3 (#4)}
\newcommand{\PRep}[4]{#1, Phys.\ Rep.\ {\bf #2}, #3 (#4)}
\newcommand{\ZPC}[4]{#1, Z.\ Phys.\ {\bf C#2}, #3 (#4)}
\newcommand{\EPJC}[4]{#1, Eur.\ Phys.\ J.\ {\bf C#2}, #3 (#4)}
\newcommand{\etal}{{\it et al.\ }}

\end{document}